\documentclass [twocolumn,aps,pra,noshowpacs]{revtex4-1}
\usepackage{epsfig,graphicx,amssymb,color,amssymb,amsmath,mathrsfs,bm}
\pdfoutput=1

\begin{document}

\title[Degenerate parametric down-conversion in nonlinear plasmonic cavity]{Degenerate parametric down-conversion facilitated by exciton-plasmon polariton states in nonlinear plasmonic cavity}


\author{Andrei Piryatinski}
\email{apiryat@lanl.gov}
\address{Theoretical Division, Los Alamos National Laboratory, Los Alamos, NM 87545, USA }

\author{Maxim Sukharev}
\address{Department of Physics, Arizona State University, Tempe, AZ 85287, USA }
\address{College of Integrative Sciences and Arts, Arizona State University, Mesa, AZ 85201, USA}

\begin{abstract}
We study the effect of degenerate parametric down-conversion (DPDC) in an ensemble of two-level quantum emitters (QEs) coupled via near-field interactions to a single surface plasmon (SP) mode of a nonlinear plasmonic cavity. For this purpose, we develop a quantum driven-dissipative model capturing non-equilibrium dynamics of the system in which incoherently pumped QEs have transition frequency tuned near the second-harmonic response of the SPs. Considering the strong coupling regime, i.e.,  the SP-QE interaction rate exceeds system dissipation rates, we find a critical SP-QE coupling attributed to the phase transition between normal and lasing steady states. Examining fluctuations above the system's steady states, we predict new elementary excitations, namely,  the exciton-plasmon polaritons formed by the two-SP quanta and single-exciton states of QEs. The contribution of two-SP quanta results in the linear scaling of the SP-QE interaction rate with the number of QEs, ${\cal N}_o$,  as opposed to the $\sqrt{{\cal N}_o}$-scaling known for the Dicke and Tavis-Cummings models. We further examine how SP-QE interaction scaling affects the polariton dispersions and power spectra in the vicinity of the critical coupling. For this purpose, we compare the calculation results assuming a finite ensemble of QEs and the model thermodynamic limit. The calculated power spectra predict an interplay of coherent photon emission by QEs near the second-harmonic frequency and correlated photon-pair emission at the fundamental frequency by the SPs (i.e., the photonic DPDC effect). 
\end{abstract}


\date{\today}
 
\maketitle


\section{Introduction}

Plasmonics has been in the spotlight in engineering,~\cite{GramotnevNatPhot:2010} chemistry,~\cite{EbbesenAcCemRes:2016,FeistASCNano:2018} and biology~\cite{HuangLaserSurgMed:2007} for more than a decade.~\cite{StockmanOptEx:2011} During this time, fascinating progress in nanofabrication provided new tools to create metal-dielectric, i.e., plasmonic, interfaces with nanometer precision.\cite{PanoiuJ.Opt:2018} Even though the surface plasmon (SP) modes in such nanostructures show relatively low characteristic quality factors, large local field enhancements~\cite{WuACSPhot:2021} make them very attractive for applications.~\cite{EbbesenAcCemRes:2016, JiangChemRev:2017} On the other hand, ultrafast optical detection techniques have offered time resolution reaching a few femtosecond limit.~\cite{VasaNatPhot:2013} These fabrication and characterization tools tremendously assist us in advancing our knowledge of the light-matter interactions at the nanoscale, where the quantum and classical properties of light and matter merge.\cite{TormaRepPrgoPhys:2014}

Despite such progress, the nonlinear plasmonics relying on the strong local field enhancement~\cite{GordonACSPhot:2018} and high nonlinear electron susceptibilities in metal~\cite{KrasavinLaserPhotRev:2018} is still in its infancy.\cite{PanoiuJ.Opt:2018} Several recent reports address nonlinear phenomena such as second/third-harmonic generation~\cite{GalantyLightSciTech:2018,WeissmanAdvOptMat:2017,YooOptExp:2019}, four-wave mixing~\cite{AlmeidaSciRep:2015,BlechmanNanoLett:2018}, and optical bistability.~\cite{WurtzPRL:2006} Moreover, the use of coherent control was reported to study the second-harmonic generation (SHG) by gold nanoparticles combining spatial localization of electromagnetic fields with laser pulse shaping.\cite{BaharPRB:2020} 

Quantum plasmonics interprets the metallic nanostructure as a plasmonic cavity where strong local electric fields enhance the spontaneous decay of organic dyes or semiconductor quantum dots representing quantum emitters (QEs).\cite{TameNatPhys:2013}  This effect is known as the plasmonic Purcell effect. Advances beyond demonstration of the Purcell effect hinge on the use of 2D plasmonic lattices facilitating delocalized hybrid plasmonic-diffraction surface lattice resonance of relatively high quality factor (low energy losses).\cite{CherquiAccChemRes:2019} Use of these structures allowed researchers to reach a strong coupling regime between QE and SP in which the SP-QE quantum exchange rate exceeds plasmonic losses and results in the formation of mixed QE and SP exciton-plasmon polariton stats.~\cite{RamezaniJOSAB:2019} Current theoretical and experimental studies of the exciton-polariton states focus on plasmonic lasing~\cite{Pusch_ACSnano:2012,Guan_ASCnano:2020,Winkler_ACSnano:2020}, polariton Bose Einstein condensation, and polariton lasing.~\cite{MartikainenPRA:2014,ZasterJPCS:2016,DeGiorgiPEP:2018,HakalaNP:2018} All these studies asume linear SP response  with the QEs providing source of nonlinearity. 

The strong SP-QE coupling regime has recently become a spotlight in nonlinear plasmonics, promising coherent photonic up/down-conversion and lasing applications. One of the authors contributed to a theoretical study demonstrating that strongly coupled nonlinear SP systems and QEs may exhibit unique second-order optical signals facilitated by mixed exciton-plasmon polariton states.\cite{DrobnyhJCP:2020,SukharevJCP:2021} Subsequent experimental studies confirmed the predictions.~\cite{LiNanoLett:2020}  In another theory report, we examined the effect of QE gain on the SHG by a plasmonic nanostructure, e.g., 2D plasmonic lattice, exhibiting nonlinear response.\cite{Sukharev_JCP:2021}  For the QEs strongly coupled with  {\em fundamental} SP mode, we found a lasing state contributing to efficient photon emission at the {\em second-harmonic} frequency. In that report, we also introduced a concept of the nonlinear plasmonic cavity in which nonlinear interaction occurs between quantized fundamental and second-harmonic plasmonic modes. Noteworthy that photonic cavities demonstrate nonlinearity due to embedded QEs only.~\cite{RibeiroPRA:2021} 

In this paper, we use the concept of nonlinear SP cavity to investigate a reverse effect of the degenerate parametric down-conversion (DPDC) of excited QEs (excitons) strongly coupled to the {\em second-harmonic} of the cavity via near-field interactions. For a localized SP mode, the effects of SHG and parametric down-conversion have been studied experimentally in light of nonlinear photon-plasmon interactions.\cite{GrossePRL:2012,HeckmannOptExp:2013}  The authors of Refs.~\cite{LootJOpt:2018,LootOptik:2018} considered spontaneous parametric downconversion in nonlinear dielectric media interfacing with metal surfaces so that the linear SP response of the metal enhances the process via {\em weak} local field coupling. None of these reports considers the effects of QEs. 

Therefore, we focus our paper on the case in which QEs forming gain medium exchange energy via the near-filed interactions with the second-harmonic of a nonlinear plasmonic cavity. As demonstrated below, the presence of QEs results in new exciton-plasmon polariton states mixing {\em two}-SP quanta with a single exciton state. Considering the  strong coupling regime, we predict a non-trivial lasing phase supporting these polariton states as elementary excitations. Our analysis of the polariton power spectra predicts coherent photon emission by QEs around the second harmonic frequency and correlated photon pairs at the fundamental frequency by the SP mode. Given the SP photon emission rate significantly exceeds the same rate for QEs, we predict the photon DPDC effect in such a system. 

We organize the paper as follows. Sec.~\ref{Sec:QuantMod} presents our quantum open system theory of the DPDC, identifying lasing steady states and providing expressions for the exciton-plasmon polariton dispersion and fluctuation auto-correlation functions. Sec.~\ref{Sec:NumDis} presents numerical calculations and discussion of the polariton dispersion relations and the auto-correlation functions. Finally, we conclude in Sec.~\ref{Sec:Conc}.

\section{Quantum driven-dissipative model}
\label{Sec:QuantMod}

\begin{figure}[b]
\begin{center}
\includegraphics[width=0.4\textwidth]{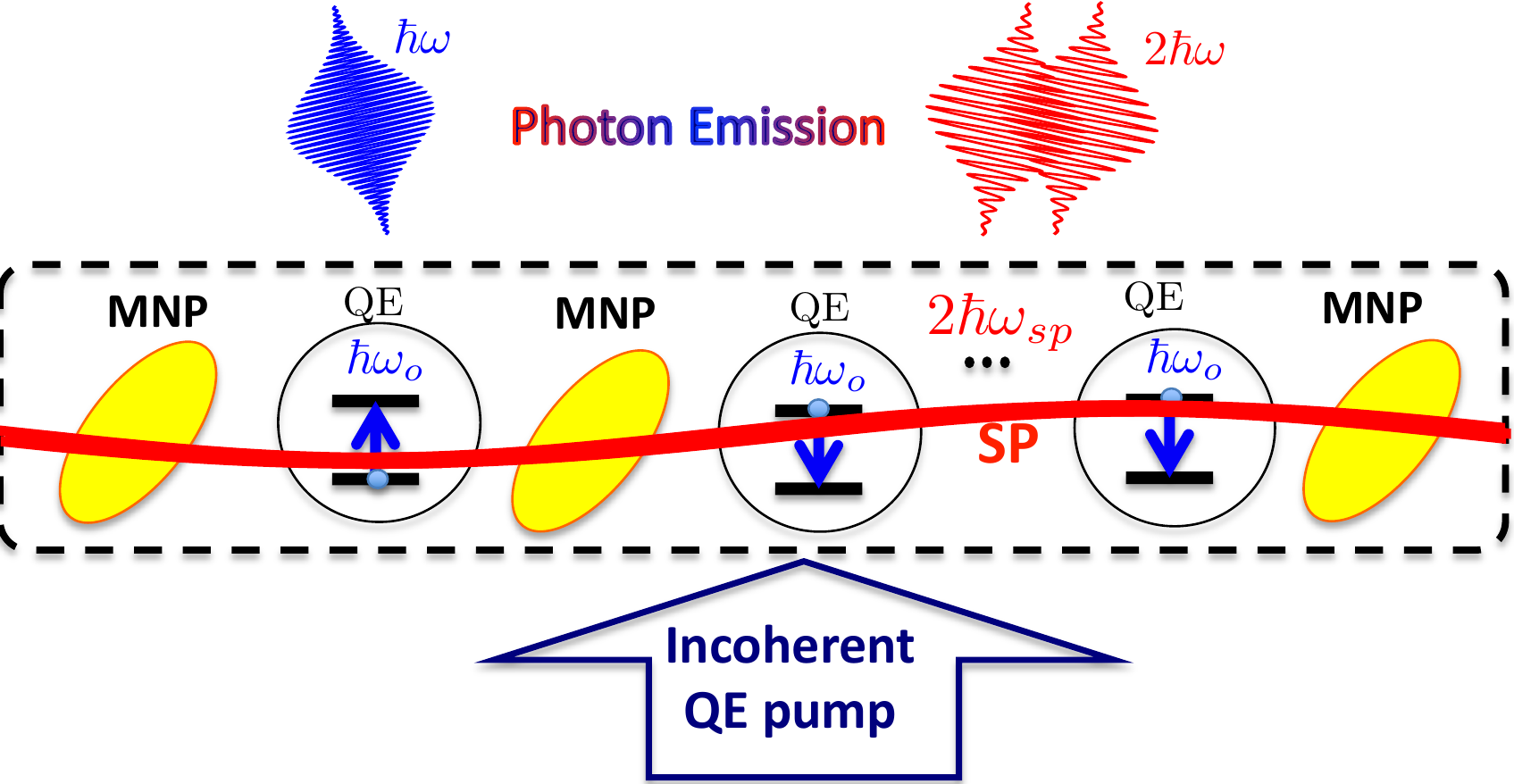}
\caption{Schematics of a plasmonic cavity formed by an ensemble of metal nanoparticles (MNPs) supporting a delocalized SP mode (red wave line). QEs whose energy $\hbar\omega_o$ is tuned near the second-harmonic response of the SP mode are coupled to the SP mode via the near-field interactions. The QEs are subject to incoherent pump providing energy for the DPDC which results in the formation of correlated two-SP states of energy $\sim\hbar\omega_{sp}$. The cavity can radiate correlated two-photon pairs  of energy $2\hbar\omega$ and  photons of energy $\hbar\omega$ due the  two-SP and QE radiative decay, respectively.}
\label{Fig:Model}
\end{center}
\end{figure}

As illustrated in Fig.~\ref{Fig:Model}, we introduce an ensemble of identical two-level QEs with the transition frequency $\omega_o$. The QEs occupy sites $n=\overline{1,{\cal N}_o}$ and interact with a delocalized SP mode whose fundamental frequency is $\omega_{sp}$. The SP mode is assumed to have second-order nonlinear response allowing for the SHG, i.e., creation of a single SP quantum with frequency $2\omega_{sp}$ or a down-conversion of a single SP quantum with frequency $2\omega_{sp}$  to two quanta each having frequency $\omega_{sp}$. By tuning the QEs around the SP second-harmonic frequency, i.e.,  $\omega_o \sim  2\omega_{sp}$, we allow for the down-conversion of the SP quanta produced by QEs. The interaction between each QE and SP is characterized by the quantum exchange rate $\lambda$ proportional to the SP second-order susceptibility $\chi_{sp}^{(2)}$.

For the introduced system, the Hamiltonian, assuming rotating wave approximation for the SP-QE interaction term, reads
\begin{eqnarray}
\label{H-def}
\hat{\cal H} &=&\hbar \omega_{sp} \hat a^\dag \hat a 
  + \hbar\omega_o\sum_{n=1}^{{\cal N}_o}\left(\hat s_n^z+\frac{1}{2}\right)
\\\nonumber&+&
    \hbar\lambda\sum_{n=1}^{{\cal N}_o}\left(\hat a^{\dag2}\hat s_n^{-}+\hat s_n^{+}\hat a^2\right).
\end{eqnarray}
Here, the SP mode is described by Bose creation (annihilation) operator $\hat a^\dag$ ($\hat a$), and the two-level QEs by spin operators $\hat s_n^\pm=\hat s_n^x \pm i\hat s^y_n$, $s_n^z$, related to the Pauli matrices as $\hat s_n^{j}=\frac{1}{2}\hat\sigma_n^{j}$ where $j=x,y,z$.  

Both the SPs and QEs are subject to energy supply originating form incoherent excitation of the QEs and dissipation processes due to environment fluctuations. Thus,  the time-evolution of such an open quantum system characterized by an operator set $\hat{\cal O}=\{\hat a, \hat s_n^{-}, \hat s_n^z\}$  can be described by the Heisenberg equation of motion 
\begin{eqnarray}
\label{HsEqMo-def}
\partial_t \hat{\cal O} &=& \frac{i}{\hbar}\left[\hat{\cal H}, \hat{\cal O} \right]
        +2\gamma_{sp} \hat{\cal L}_{\hat a}[\hat {\cal O}]
	+\gamma_\uparrow\sum\limits_{n=1}^{{\cal N}_o} \hat{\cal L}_{\hat s^{+}_n}[\hat {\cal O}]
\\\nonumber &+&	
	\gamma_\downarrow\sum\limits_{n=1}^{{\cal N}_o}\hat{\cal L}_{\hat s^{-}_n}[\hat {\cal O}]
	+ \gamma_\phi\sum\limits_{n=1}^{{\cal N}_o}\hat{\cal L}_{\hat s_n^z}[\hat {\cal O}],
\end{eqnarray}
where the first term in the right-hand side accounts for the coherent dynamics governed by the Hamiltonian $\hat {\cal H}$. The second term describes SP decay with the rate $2\gamma_{sp}$ and the last three terms account for incoherent pumping of QEs with the rate $\gamma_\uparrow$, QE decay with the rate $\gamma_\downarrow$ and pure dephasing with the rate $\gamma_\phi$. Here, the Lindblad operator describing dissipation process in the Markovian approximation is
\begin{eqnarray}
\label{LdOp-def}
\hat{\cal L}_{\hat C}[\hat {\cal O}] &=& \frac{1}{2}\left(\hat C^\dag\left[ \hat {\cal O},  \hat C \right] 
				+\left[ \hat C^\dag ,\hat {\cal O}\right]  \hat C\right).
\end{eqnarray}

The substitution of the Hamiltonian~(\ref{H-def}) and  $\hat{\cal O}=\{\hat a, \hat s_n^{-}, \hat s_n^z\}$ into Eqs.~(\ref{HsEqMo-def}) -- (\ref{LdOp-def}) results in the following set of coupled Heisenberg equations of motion for the SP and QE degrees of freedom
\begin{eqnarray}
\label{a-OpEqMo}
&~&\partial_t \hat a ~= -\left(i \omega_{sp} +\gamma_{sp} \right)\hat a - 2i\lambda \hat a^\dag \sum_{n=1}^{{\cal N}_o} \hat s_n^{-},
\\\label{sm-OpEqMo}
&~&\partial_t s_n^{-} = -\left(i \omega_o +\gamma_o \right)s_n^{-}+2i\lambda s_n^z \hat a^2,
\\\label{sz-OpEqMo}
&~&\partial_t s_n^z = -\gamma_t\left(s_n^z -\frac{d_o}{2}\right)+i\lambda\left(\hat a^{\dag2} s_n^{-}-s_n^{+}\hat a^2 \right).
\end{eqnarray}
Eqs.~(\ref{sm-OpEqMo}) and (\ref{sz-OpEqMo}) contain pumping rate dependent QE dephasing,  $\gamma_o=\gamma_\downarrow/2+\gamma_\uparrow/2+\gamma_\phi$ and  population decay, $\gamma_t=\gamma_\uparrow+\gamma_\downarrow$, rates. Eq.~(\ref{sz-OpEqMo}) also contains the population inversion parameter
\begin{eqnarray}
\label{do-def}
d_o=\frac{\gamma_\uparrow-\gamma_\downarrow}{\gamma_\uparrow+\gamma_\downarrow},
\end{eqnarray}
whose value varies in the interval $-1\leq d_o<0$ ($0<d_o\leq 1$) for $\gamma_\uparrow<\gamma_\downarrow$ ($\gamma_\uparrow>\gamma_\downarrow$) indicating that QEs are pumped below (above) population inversion threshold. 

Eq.~(\ref{a-OpEqMo}) has a well familial form of  equation of motion for the parametric down-conversion processes~\cite{Loudon_book:10}. While conventional DPDC occurs due to a coherent optical pump tuned in resonance with the second-harmonic, our model replaces the laser pump with the QE coherences interacting with the counter-rotating SP coherence in the last term of Eq.~(\ref{a-OpEqMo}).  

\subsection{Lasing phase transition}
\label{Sec-ststs}

To get a better insight into this model, we first identify its nontrivial steady state appearing in the strong coupling regime. We start the analysis by deriving mean-field equations of motion for uncorrelated SP and QE degrees of freedom. Specifically, for the two-SP coherence, $\langle\hat a^2\rangle$, and SP population, $\langle\hat a^\dag \hat a\rangle$, and the following normalized coherence $\langle\hat s_{-}\rangle=\sum_{n=1}^{{\cal N}_o}\langle\hat s_n^{-}\rangle/{\cal N}_o$ and population inversion $\langle\hat s_{z}\rangle=\sum_{n=1}^{{\cal N}_o}\langle\hat s_n^{z}\rangle/{\cal N}_o$ of QEs. To find conditions for the lasing phase transition, we break U(1) gauge symmetry by transforming SP and QE coherences into rotating frame, i.e., replacing $\hat a\rightarrow\hat a e^{-i\omega_{L}t/2}$ and $\hat s_{-}\rightarrow\hat s_{-} e^{-i\omega_{L}t}$. In these variables  Eqs.~(\ref{a-OpEqMo})--(\ref{sz-OpEqMo}), result in a set of equations of motion
\begin{eqnarray}
\label{a2MfEqMo}
\partial_t \langle\hat a^2\rangle &=& -2\left(i \delta\omega_{sp} +\gamma_{sp} \right)\langle\hat a^2\rangle 
\\\nonumber &~&
- 2i{\cal N}_o\lambda\left(2\langle\hat a^\dag\hat a\rangle+1\right) \langle \hat s_{-}\rangle,
\\\label{sm-MfEqMo}
\partial_t \langle \hat s_{-}\rangle &=& -\left(i \delta\omega_o +\gamma_o \right)\langle\hat s_{-} \rangle
            +2i\lambda \langle\hat s_z\rangle \langle\hat a^2\rangle,
\\\label{nph-MfEqMo}
\partial_t \langle\hat a^\dag\hat a\rangle &=& -2 \gamma_{sp}\langle\hat a^\dag\hat a\rangle
\\\nonumber&~&
+ 2i\lambda{\cal N}_o\left[\langle\hat s_{+}\rangle\langle\hat a^2\rangle
    -\langle\hat a^{\dag 2}\rangle \langle\hat s_{-}\rangle \right]
\\\label{sz-MfEqMo}
\partial_t \langle\hat s_z\rangle &=& -\gamma_t\left(\langle \hat s_z\rangle -\frac{d_o}{2}\right)
\\\nonumber&~&
                -i\lambda\left[\langle\hat s_{+}\rangle\langle\hat a^2\rangle
    -\langle\hat a^{\dag 2}\rangle \langle\hat s_{-}\rangle \right],
\end{eqnarray}
with the short hand notations $\delta\omega_{sp}=\omega_{sp}-\omega_L/2$, and $\delta\omega_o=\omega_o-\omega_L$ for the frequency offsets due to the lasing frequency, $\omega_L$, determined below. 

Denoting the steady states of the two-SP coherence as $\bar\alpha_2$, SP population as $\bar n_{sp}$, QE coherence as $\bar s_{-}$, and population inversion as $\bar s_z$, we present their non-trivial expressions  
\begin{eqnarray}
\label{a2-StSo}
&~&|\bar\alpha_2|=\bar n_{sp},
\\\label{nsp-StSo}
&~&\bar n_{sp}= {\cal N}_o
    \frac{d_o\gamma_t}{4\gamma_{sp}}
    \left (1 \pm\sqrt{1-\left[\frac{\lambda_c}{{\cal N}_o\lambda}\right]^2}\right),
\\\label{sz-StSo}
&~&\bar s_{z} = \frac{d_o}{4}
    \left(1 \mp \sqrt{1-\left[\frac{\lambda_c}{{\cal N}_o\lambda}\right]^2} \right),
\\\label{sm-StSo}
&~&|\bar s_{-}|^2 = \frac{\delta\omega_{sp}^2+\gamma_{sp}^2}{\left(2{\cal N}_o\lambda\right)^2},
\end{eqnarray}
satisfying Eqs.~(\ref{a2MfEqMo})--(\ref{sz-MfEqMo}) with the time derivatives set to zero. This solution exists in the QE inversion regime, i.e., $d_o>0$, and above the SP-QE critical coupling rate 
\begin{eqnarray}
\label{lmbd_c-def}
\lambda_c^2 =  \frac{4\gamma_{sp}^2}{d_o^2\gamma_o\gamma_t} 
    \left( \delta\omega_o^2 +\gamma_o^2\right),
\end{eqnarray}
i.e., ${\cal N}_o\lambda>\lambda_c$. Associated lasing frequency offsets are
\begin{eqnarray}
\label{dwo-def}
\delta\omega_o &=& \frac{\omega_o-2\omega_{sp}}{1+2\gamma_{sp}/\gamma_o},
\\\label{dwsp-def}
\delta\omega_{sp} &=& \frac{\omega_{sp}-\omega_o/2}{1+\gamma_o/(2\gamma_{sp})}.
\end{eqnarray}
Below the critical coupling Eqs.~(\ref{a2MfEqMo})--(\ref{sz-MfEqMo})  have a trivial steady state with $\bar s_z=d_o/2$, $\bar\alpha_2=\bar n_{sp}=\bar s_{-} = 0$ and $\delta\omega_o=\delta\omega_{sp}=0$. 

The dependence of our critical coupling, $\lambda_c$, on the dissipation rates (Eq.~(\ref{lmbd_c-def})) shows that for ${\cal N}_o\lambda > \lambda_c$ our system enters the strong coupling regime, i.e., the interaction rate exceeds any dissipation rate. In the case of Dicke and Tavis-Cummings models, strong coupling regime can be achieved by increasing the number of QEs, ${\cal N}_o$, since, effective coupling rate, $\sqrt{{\cal N}_o}\lambda$, scales as $\sqrt{{\cal N}_o}$.\cite{Kirton_AdvQT:2019,Piryatinski_PRR:2020,Sukharev_JCP:2021}  Interestingly in our model, the effective coupling, ${\cal N}_o\lambda$, scales linearly with ${\cal N}_o$ as can be noticed in  Eqs.~(\ref{nsp-StSo}) and (\ref{sz-StSo}). This allows one to reach strong coupling regime for a smaller number of QEs compared to the Dicke and Tavis-Cummings models. The difference in the scaling reflects the fact that in our model single energy quantum due to a QE is coupled to two-SP quanta whereas in the aforementioned models a single cavity quantum couples a single QE excitation. 

The mean field steady states (Eqs.~(\ref{nsp-StSo})--(\ref{sm-StSo})) are exact in the thermodynamic limit difined as ${\cal N}_o\rightarrow \infty$ and $\lambda\rightarrow 0$ preserving a finite value of their product ${\cal N}_o\lambda$. As we demonstrate in Sec.~\ref{Sec:NumDis}, obtained steady state solution can be used as a good approximation to examine formation of the lasing phase in finite but large enough ensembles of QEs. Much better description of the finite size ensemble can be obtained by means of equations of motion which  account for the correlations between QEs and SP degrees of freedom.~\cite{Kirton_AdvQT:2019,Piryatinski_PRR:2020,Kirton_NJP:2018} Thus, we introduce correlated SP-QE coherence $c_{+sp}=\sum\limits_{n}\langle\hat s^{+}_n \hat a^2\rangle/{\cal N}_o$, SP-SP long range coherence $c_{+-}=\sum\limits_{n\neq m}\langle\hat s^{+}_n \hat s^{-}_m\rangle/[{\cal N}_o({\cal N}_o-1)]$, and keep  $n_{sp}=\langle\hat a^\dag \hat a\rangle$,  and $s_{z}=\sum_{n}\langle\hat s_n^{z}\rangle/{\cal N}_o$, to form a new set of variables. 

Following the methodology used in ~\cite{Kirton_AdvQT:2019,Piryatinski_PRR:2020,Kirton_NJP:2018} , we derived the following closed set of equations of motion for these variables
\begin{eqnarray}
\label{spphi-EqMo}
\partial_t c_{+sp} &=& -\left[i\left(2\omega_{sp}-\omega_o\right)+2\gamma_{sp}+\gamma_o\right]c_{+sp} 
\\\nonumber &~&
    -2i\lambda\left[2n^2_{sp} s_z  +\left(2 n_{sp} +1\right)
\right.\\\nonumber&~&\left.\times
    \left(s_z+1/2+({\cal N}_o-1)c_{+-}\right)\right],
\\\label{spsm-EqMo}
\partial_t c_{+-} &=& -2\gamma_o c_{+-} + 2i\lambda s_z\left(c_{+sp}-c^*_{+sp} \right),
\\\label{sz-EqMo}
\partial_t s_z &=& -\gamma_t\left(s_z -d_o/2\right)-i\lambda\left(c_{+sp}-c^*_{+sp} \right),
\\\label{nph-EqMo}
\partial_t n_{sp} &=& -2 \gamma_{sp}n_{sp} + 2i\lambda{\cal N}_o\left(c_{+sp}-c^*_{+sp} \right),
\end{eqnarray}
starting with the operator Eqs.~(\ref{a2MfEqMo})--(\ref{sz-MfEqMo}).  Steady states of Eqs.~(\ref{spphi-EqMo})--(\ref{nph-EqMo}) will be numerically identified and results will be compared with the mean-filed theory in Sec.~\ref{Sec:NumDis}.

\subsection{Cavity polaritons and related auto-correlation functions}

Now, we turn our attention to the two-SP, ${\alpha}_2= \langle\hat a^2\rangle-\bar\alpha_2$, and QEs, $ s_{-}=\langle\hat s_{-}\rangle-\bar s_{-}$ coherence fluctuations. According to Eqs.~(\ref{a2MfEqMo}) and (\ref{sm-MfEqMo}) the fluctuations  satisfy linearized equation of motion
\begin{eqnarray}
\label{a2sm-flEqMo}
\partial_t {\bm\xi}(t)= {\cal M} {\bm\xi}(t). 
\end{eqnarray}
where, the fluctuation vector is ${\bm \xi}=({\alpha}_2, s_{-})^T$ and stability matrix
\begin{eqnarray}
\label{Mst}
{\cal M} = \begin{bmatrix}
-2\left(i \delta\omega_{sp} +\gamma_{sp} \right) & - 2i{\cal N}_o\lambda(2\bar n_{sp}+1) \\
2i\lambda \bar s_z & -i \delta\omega_o -\gamma_o  
\end{bmatrix}, 
\end{eqnarray}
depends on the steady state parameters $\bar s_z$, $\bar{n}_{sp}$, $\delta\omega_o$, $\delta\omega_{sp}$. The time-dependent and Fourier transformed solutions of linear Eq.~(\ref{a2sm-flEqMo}) in terms of the initial condition vector, ${\bm \xi}(0)$, are given in Appendix~\ref{Appx:I}. As expected the solution depends on the eigenvalues of ${\cal M}$,
\begin{eqnarray}
\label{Lmbdpm}
&~&\Lambda_\pm = i\delta\omega_{sp}+\gamma_{sp} +\frac{i\delta\omega_o+\gamma_o}{2} 
\\\nonumber&~&\pm
\left\{\left[i\omega_{sp}+\gamma_{sp} -\frac{i\omega_o+\gamma_o}{2} \right]^2
+ 4{\cal N}_o \lambda^2 \bar s_z [2\bar n_{sp}+1]\right\}^{1/2},
\end{eqnarray}
as a function of the SP-QE coupling rate. 

\begin{figure*}[t]
\begin{center}
\includegraphics[width=1.0\textwidth]{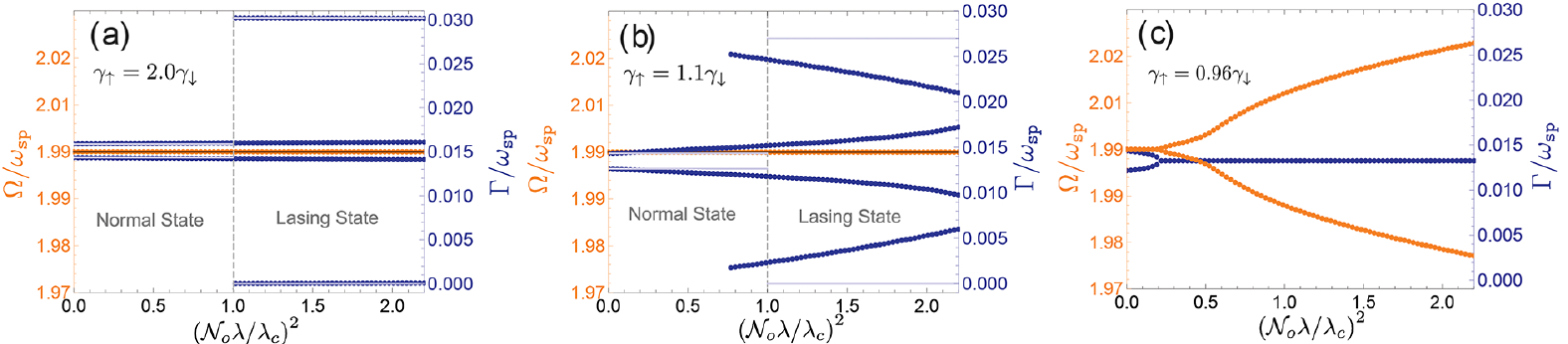}
 \caption{Calculated polariton dispersion curves as a function of normalized QE-SP effective coupling rate for various values of the incoherent pumping rates $\gamma_\uparrow$ indicated in each panel: (a) and (b) QE are pumped above population inversion with the inversion ratios (Eq.~(\ref{do-def})) of $d_o=0.33$ and  $d_o=0.046$, respectively. (c) QEs are pumped slightly below population inversion with $d_o=-0.02$.}   
\label{Fig:PolDis}
\end{center}
\end{figure*}

Next we introduce, the following auto-correlation functions associated with the two-SP and QE coherences
\begin{eqnarray}
\label{Csp-def}
C_{sp}(t) &=& \left\langle\hat a^{\dag 2}(t)\hat a^{2}(0)\right\rangle,
\\\label{Cqe-def}
C_{qe}(t) &=&  \left\langle\hat s_{+}(t)\hat s_{-}(0)\right\rangle,
\end{eqnarray}
respectively. To account for the fluctuation dynamics in terms of these correlation functions, we use the quantum regression theorem.\cite{Loudon_book:10,Piryatinski_PRR:2020} According to this theorem, $C_{sp}(t)$ complimented by the cross-correlation function $C_{X1}=\left\langle\hat s_{+}(t)\hat a^{2}(0)\right\rangle$ form a vector ${\bm \xi}^*(t)=\left(C_{sp}(t),C_{X1}(t)\right)^T$ that satisfies complex conjugate Eq.~(\ref{a2sm-flEqMo}). The same applies to the pair of $C_{qe}(t)$ and $C_{X2}=\left\langle\hat a^{\dag 2}(t)\hat s_{-}(0)\right\rangle$ in the vector representation  ${\bm \xi}^*(t)=\left(C_{X2}(t),C_{qe}(t)\right)^T$. Using Fourier transformed solution of Eq.~(\ref{a2sm-flEqMo}) given in Appendix~\ref{Appx:I}, we provide explicit form for the frequency domain representation of the auto-correlation functions
\begin{eqnarray}
\label{Csp-w}
C_{sp}(\omega) &=& \frac{i\left(\omega-\omega_o\right) +\gamma_o}
                    {\left(i\omega+\Lambda^*_{+}\right)\left(i\omega+\Lambda^*_{-}\right)}
                    \left\langle\hat a^{\dag 2}\hat a^{2}\right\rangle
\\\nonumber &+& 
        \frac{2i{\cal N}_o\lambda \left(2\bar n_{sp}+1\right)}
                {\left(i\omega+\Lambda^*_{+}\right)\left(i\omega+\Lambda^*_{-}\right)}
    \left\langle\hat s_{+}\hat a^{2}\right\rangle,
\\\label{Cqe-w}
C_{qe}(\omega) &=&  \frac{2i\lambda \bar s_{z}}
        {\left(i\omega+\Lambda^*_{+}\right)\left(i\omega+\Lambda^*_{-}\right)}
 \left\langle\hat a^{\dag 2}\hat s_{-}\right\rangle 
\\\nonumber &+&
\frac{i\left(\omega-2\omega_{sp}\right)+2\gamma_{sp}}
                   {\left(i\omega+\Lambda^*_{+}\right)\left(i\omega+\Lambda^*_{-}\right)}
        \left\langle\hat s_{+}\hat s_{-}\right\rangle.
\end{eqnarray}
To extract the fluctuation dynamics, zero-time operator correlation functions entering into Eqs.~(\ref{Csp-w}) and (\ref{Cqe-w}) as initial conditions should be evaluated in the steady state. Adopting the the mean-field approximation, we factorize the mean operator products as 
\begin{eqnarray}
\label{ic11-mf}
\left\langle\hat a^{\dag 2}\hat a^{2}\right\rangle &=& 2\bar n_{sp}^2,
\\\label{ic12-mf}
\left\langle\hat a^{\dag 2}\hat s_{-}\right\rangle &=& 
\left\langle\hat s_{+}\hat a^{2}\right\rangle 
=\bar n_{sp} |\bar s_{-}| e^{-i\varphi_o},
\\\label{ic22-mf}
\left\langle\hat s_{+}\hat s_{-}\right\rangle &=& |\bar s_{-}|^2 + \bar s_z + 1/2.
\end{eqnarray}
In Eq.~(\ref{ic12-mf}), $\varphi_o$  is initial phase of the QE spontaneous coherence, $s_{-}$. 

Accordingly below critical coupling, the two-SP auto-correlation function vanishes, since, no coherent energy exchange between SP-QE exists. In contrast, the QE auto-correlation function acquires a finite but trivial form 
\begin{eqnarray}
\label{Cqe0}
C_{qe}(\omega) &=& 
\frac{d_o+1/2}{i\left(\omega-\omega_o\right)+\gamma_o},
\end{eqnarray}
whose real part describes photon emission lineshape for the non-interacting QEs. Non-vanishing prefactor $d_o+1/2$ reflects  incoherent pumping resulting in the energy emission. Above critical coupling, full set of Eqs.~(\ref{Csp-w}) and (\ref{Cqe-w}) should be used to evaluate the correlation functions. It should be complimented with Eq.~(\ref{Lmbdpm}) for polariton branches and Eqs.~(\ref{nsp-StSo})--(\ref{dwsp-def}) for the lasing steady state parameters.

As stated above, the steady states of our model can also be calculated using a set of Eq.~(\ref{spphi-EqMo})-(\ref{nph-EqMo}) directly accounting for the QE-SP correlations. In this case, the zero-time correlation functions entering Eqs.~(\ref{Csp-w}) and (\ref{Cqe-w}) can be factorized as   
\begin{eqnarray}
\label{ic11-cr}
\left\langle\hat a^{\dag 2}\hat a^{2}\right\rangle &=& 2\bar n_{sp}^2,
\\\label{ic12-cr}
\left\langle\hat s_{+}\hat a^{2}\right\rangle &=& \left\langle\hat a^{\dag 2}\hat s_{-}\right\rangle^*=\bar{c}_{+sp},
\\\label{ic22-cr}
\left\langle\hat s_{+}\hat s_{-}\right\rangle &=& \bar{c}_{+-}+ \bar s_z + 1/2.
\end{eqnarray}
Here, the right-hand side contains the steady state SP population, $\bar n_{sp}$, and QE inversion, $\bar s_z$, as well as the steady state correlated two-SP-QE coherence, $\bar{c}_{+sp}$,  and the long range QE-QE coherences, $\bar{c}_{+-}$.  

\section{Numerical Results and Discussion}
\label{Sec:NumDis}

In this section, we calculate the polariton dispersion (Eq.~(\ref{Lmbdpm})) as a function of the SP-QE coupling and power spectra of the SP and QE elementary excitations. Let us  define the spectra as $S_{sp}=\text{Re}[C_{sp}(\omega)]/{\cal N}_o^2$ and $S_{qe}= \text{Re}[C_{qe}(\omega)]$ using Eqs.~(\ref{Csp-def}) and  (\ref{Cqe-def}) for the auto-correlation functions. The correlation functions are calculated using both the mean-field (Eq.~(\ref{a2-StSo})--(\ref{dwsp-def})) and correlated QE-SP (Eqs.~(\ref{spphi-EqMo})--(\ref{nph-EqMo}))  models allowing for comparison between their predictions.  Parameterization of the models involves setting SP and QE energies to $\hbar\omega_{sp}=\hbar\omega_o/2=1.95\text{eV}$, QE decay (dephasing) rate to $\hbar\gamma_\downarrow=0.014\text{eV}$ ($\hbar\gamma_\phi=0.01\text{eV}$), and the SP dephasing rate to $\hbar\gamma_{sp}=0.014\text{eV}$. These parameters are consistent with our previous calculations reported in Ref.~\cite{Sukharev_JCP:2021}. Finally, for the correlated SP-QE model, Eqs.~(\ref{spphi-EqMo})--(\ref{nph-EqMo}) were solved numerically by setting the number of QEs to ${\cal N}_o=5000$. 

Figure~\ref{Fig:PolDis} shows the dispersion of polariton frequency $\Omega = \text{Im}[\Lambda_\pm]$ (orange curves) and dissipation rate $\Gamma = \text{Re}[\Lambda_\pm]$ (blue curves) as a function of normalized coupling parameter. Probing various values of the incoherent pumping rate and associated (Eq.~(\ref{do-def})) inversion parameter, $d_o$, reveals several distinct features of polariton states and limitations of the mean-field model. For a relatively high pumping rate, $d_o=0.33$, results of both mean-field and correlated SP-QE calculations are the same (see panel~a). In this case, the lower and upper polariton frequencies do not split as the system transitions from the normal (i.e., trivial ) steady state to the lasing state. In contrast,  the polariton dissipation rate branches split above the critical coupling, and the lower branch reaches zero. This trend is due to the QE gain, which fully compensates for the losses indicating lasing regime. For a lower inversion characterized by $d_o=0.46$, panel~b shows no changes in the behavior of the mean-field solution (thin blue lines). However, the dissipation rate calculated using the correlated QE-SP model (blue dots) behaves differently. Specifically, the polariton frequency splitting associated with the strong coupling regime occurs below the {\em mean-field}  threshold ($\sim 0.75\lambda_c$). At the same time, the lower polariton dissipation rate never reaches zero, i.e., the QE gain does not fully compensate the losses. 

\begin{figure*}[t]
\begin{center}
\includegraphics[width=0.8\textwidth]{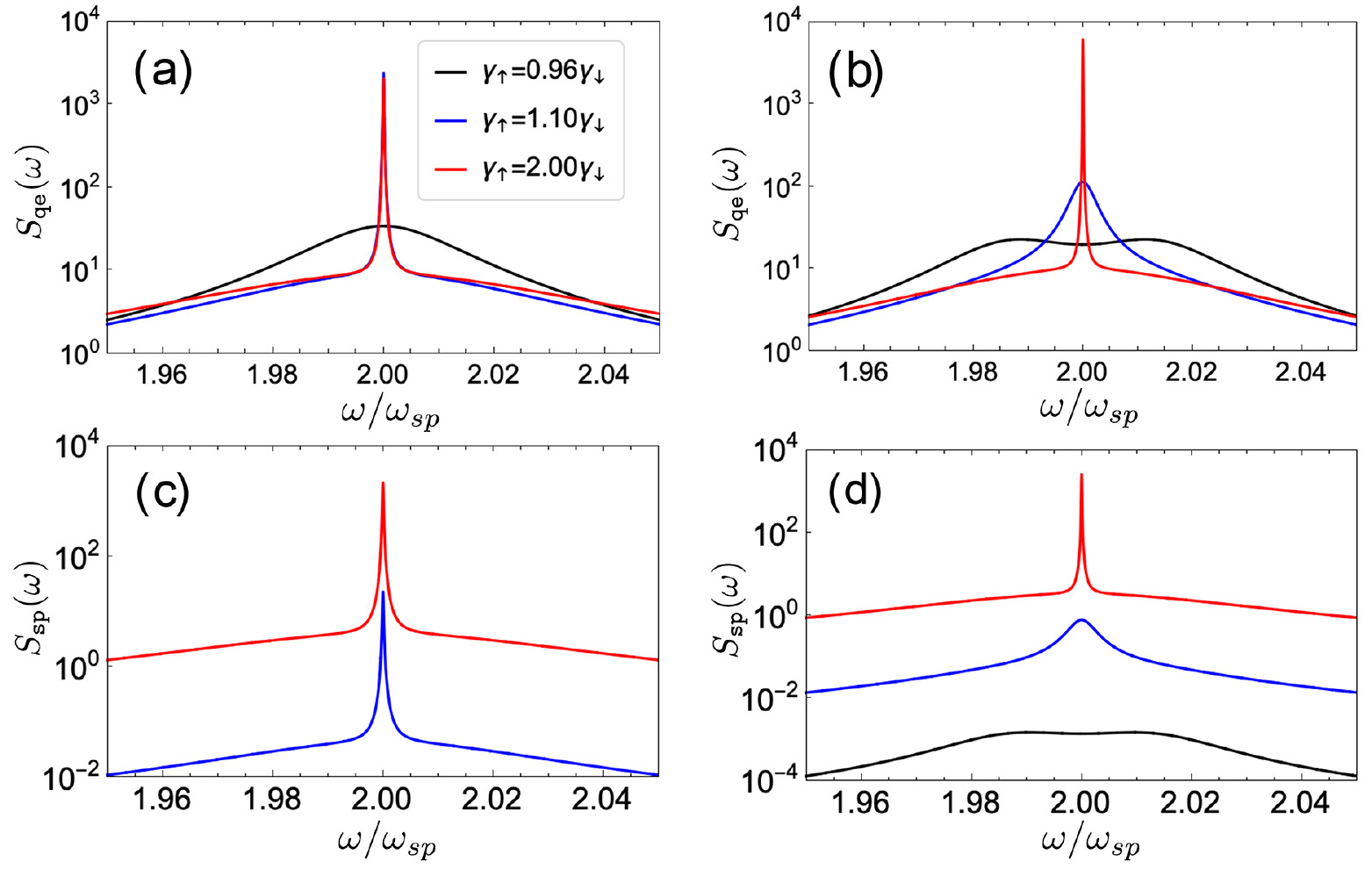}
\caption{Calculated power spectra for (a), (b) ensemble of QEs normalized per ${\cal N}_o$ and (c), (d) correlated two-SP states normalized per ${\cal N}_o^2$ in strong couplings regime characterized by $\lambda=2.1\lambda_c$ and values of incoherent pumping rate $\gamma_\uparrow$ same as in Fig.~\ref{Fig:PolDis}. The mean-field calculations are presented in the left column and the correlated QE-SP model is used in the right column.}   
\label{Fig:Sw}
\end{center}
\end{figure*}

These features indicate deviations in the properties of finite-size QE ensembles from the predictions of the mean-field theory. The latter, indeed, is exact only in the thermodynamic limit. For example, failure of the lower polariton dissipation branch to reach zero is a signature of cross-over behavior between normal and lasing states rather than the lasing phase transition. Furthermore, below the QE inversion (panel~c), our mean-field theory is not defined, but the correlated QE-SP calculations produce a satisfactory result. They show a well-known behavior of the polariton branches while crossing a threshold between weak and strong coupling regimes at ${\cal N}_o\lambda\approx 0.2\lambda_c$. At this point, the polariton frequency (orange dots) experiences the Rabi splitting, and the QE and SP dissipation rates (blue dots) merge into a single rate. The trend of lowering polariton strong coupling threshold, observed in panels~b and c, with the pumping rate decrease, is likely related to the dependence of the dissipation rates on the pumping rate (see comments following Eqs.~(\ref{a-OpEqMo}) and (\ref{sz-OpEqMo})).          

Trends in the variations of the QE power spectra, calculated in the mean-field approximation (Fig.~\ref{Fig:Sw}a) and using the QE-SP correlated model (Fig.~\ref{Fig:Sw}b), are consistent with the discussed trends in polariton dispersion (Fig.~\ref{Fig:PolDis}). Calculated above critical coupling at ${\cal N}_o\lambda = 2.0\lambda_c$, both panels of the plot show narrow lasing peaks at the pumping rate twice the decay rate (red curves). Differences appear when the pumping rate gets closer to the inversion threshold. In this situation (blue lines), the mean-field spectrum still has a sharp lasing peak, whereas the correlated QE-SP model predicts a broader peak. The broadening occurs because the lower polariton dissipative branch (blue dots in Fig.~\ref{Fig:PolDis}b) does not reach zero, i.e., the QE gain does not fully compensate for the losses.    

Below the inversion threshold, our mean-field model predicts a single broad peak (black curve in panel~a) originating from the trivial limit of the fluctuation correlation function (Eq.~(\ref{Cqe0})) with the resonance at $\omega_o$. In contrast, the correlated QE-SP model (black line in panel~b) predicts two resonances due to the Rabi splitting of the polariton frequency (orange dots in Fig.~\ref{Fig:PolDis}c). The two-SP power spectra in panels~c and d of  Fig.~\ref{Fig:Sw} show the same features as panels~a and b. The mean-field model predicts no DPDC effect below inversion (notice the absence of a black curve in pane~c). To understand this, recall that the two-SP coherence directly couples to the QE steady state coherence $\bar s_{-}$ (see, e.g., Eq.~(\ref{H-def})). The latter vanishes below the population inversion threshold and cancels the two-SP coherence.      

As illustrated in Fig.~\ref{Fig:Model}, the radiative decay of QEs and two-SP states results in the photon emission around the second-harmonic frequency and correlated fundamental-frequency photon-pair production, respectively. The QE and SP spectra are appropriately normalized, so comparing peak values, we see an even quantum number distribution between the QEs and two-SP degrees of freedom. In contrast, the number of photons produced via the radiative decay of QEs and two-SP states depends on their material-specific spontaneous emission rates. This quantity in plasmonic nanostructures can be orders of magnitude higher than in organic chromophores or semiconductor QEs. As a result, we expext that the photon emission spectra should be dominated by the correlated two-photon pairs resulting in the photon DPDC effect.

\section{Concluding Remarks}
\label{Sec:Conc}

We have examined the mechanism of DPDC in nonlinear plasmonic cavities in which the incoherently pumped QEs interact with the SP second-harmonic mode via near-field coupling. As demonstrated, such a model allows for a lasing phase transition in the strong coupling regime. Interestingly, the effective coupling rate in our model scales as ${\cal N}_o\lambda$ in contrast to conventional scaling of $\sqrt{{\cal N}_o}\lambda$ known for the Dicke and Tavis-Cummings models facilitating the strong coupling regime for smaller ensembles of QEs. Comparing the finite-size ensemble of QEs with the model thermodynamic limit, we have found that in the former case, the strong coupling threshold lowers as the QE pumping rate decreases. Furthermore, the finite-size ensemble of QEs pumped below population inversion is predicted to demonstrate the polariton Rabi splitting and the DPDC effect. In the thermodynamic limit, no such effects take place. Our model also predicts that the lasing threshold in finite-size systems occurs for high pumping rate than in the thermodynamic limit. 

We expect that 2D plasmonic lattices showing the lattice-plasmon resonances~\cite{RamezaniJOSAB:2019,Sukharev_JCP:2021} could be a good material platform for observation predicted DPDC effect. However, our basic quantum optics model cannot incorporate some material-specific features of the lattice-plasmon resonances. This calls for numerical simulations~\cite{SukharevJPCM:2017}  accounting for the composition and geometry of metal nanoparticles, detailed mechanisms of nonlinearity, and QE electronic structure. In particular, the simulations should address the question of photon emission yield for the correlated two-SP states compared to the emission yield of the QEs. Finally, reported progress on developing new plasmonic nanomaterials, e.g.,  doped semiconductors,~\cite{AgrawalChenRev:2018,Dolgopolova_NanoHorizons:2022} with SP resonances in the near-infrared, will pave the way for practical applications of the proposed effect in on-chip photonic devices.

\section{Acknowledgements}
\label{Sec:Acknowledgements}

This work was performed, in part, at the Center for Integrated Nanotechnologies, an Office of Science User Facility operated for the U.S. Department of Energy (DOE) Office of Science by Los Alamos National Laboratory (Contract 89233218CNA000001) and Sandia National Laboratories (Contract DE-NA-0003525) (user project No. 2021BC0087). A.P. would like to acknowledge support via LDRD exploratory research grant No. 20200407ER. M.S. is grateful for the financial support by the Air Force Office of Scientific Research under Grant No. FA9550-22-1-0175. We would like to thank Professor Joseph Zyss for stimulating discussions.   

\appendix

\section{Solution of Eqs.~(\ref{a2sm-flEqMo}) and (\ref{Mst})}
\label{Appx:I}

Solution of Eqs.~(\ref{a2sm-flEqMo}) and (\ref{Mst}) is
\begin{eqnarray}
\label{xioft}
\bm \xi(t) &=&  {\cal G}(t)\bm \xi(0)
\end{eqnarray}
where $\bm\xi(0)$ is the vector of initial conditions and the Green function, ${\cal G}(t)$,  matrix elements are
\begin{eqnarray}
\label{g11-t}
g_{11}(t)&=&\frac{1}{2}\left[ 1+\frac{i\omega_o+\gamma_o-2\left(i\omega_{sp}+\gamma_{sp}\right)}
                    {\Lambda_{+}-\Lambda_{-}}\right] e^{-\Lambda_{-}t}\;\;\;\;
\\\nonumber &+&
            \frac{1}{2}\left[ 1-\frac{i\omega_o+\gamma_o-2\left(i\omega_{sp}+\gamma_{sp}\right)}
                    {\Lambda_{+}-\Lambda_{-}}\right]e^{-\Lambda_{+}t},
\\\label{g12-t}
g_{12}(t) &=& \frac{2i{\cal N}_o\lambda \left(2\bar n_{sp}+1\right)}{\Lambda_{+}-\Lambda_{-}}
                    \left(e^{-\Lambda_{+}t}-e^{-\Lambda_{-}t}\right),
\\\label{g21-t}
g_{21}(t) &=& \frac{2i\lambda \bar s_{z}}{\Lambda_{+}-\Lambda_{-}}
                    \left(e^{-\Lambda_{-}t}-e^{-\Lambda_{+}t}\right),
\\\label{g22-t}            
g_{22}(t)&=& \frac{1}{2}\left[ 1+\frac{i\omega_o+\gamma_o-2\left(i\omega_{sp}+\gamma_{sp}\right)}
                    {\Lambda_{+}-\Lambda_{-}}\right] e^{-\Lambda_{+}t}
\\\nonumber &+&
            \frac{1}{2}\left[ 1-\frac{i\omega_o+\gamma_o-2\left(i\omega_{sp}+\gamma_{sp}\right)}
                    {\Lambda_{+}-\Lambda_{-}}\right]e^{-\Lambda_{-}t}.
\end{eqnarray}
where, $\Lambda_\pm$ given by Eq.~(\ref{Mst}).

Next, we perform the Fourier transform of the Green function
\begin{eqnarray}
\label{xiofw}
{\cal G}(\omega) = \int\limits_{-\infty}^\infty {\cal G}(t) e^{i\delta\omega t} dt
\end{eqnarray}
where $\delta\omega=\omega-\omega_L$ and the lower integration limit assumes that ${\cal G}(-|t|)=0$. This allows us to represent the spectra in terms of the initial condition vector $\xi(0)$.  The result of the calculation is
\begin{eqnarray}
\label{xiofttt}
\bm \xi(\omega) &=&  {\cal G}(\omega)\bm \xi(0),
\end{eqnarray}
with the Green function matrix elements
\begin{eqnarray}
\label{g11-w}
g_{11}(\omega)&=&\frac{i\left(\omega-\omega_o+i\gamma_o\right)}
                    {\left(\omega+i\Lambda_{+}\right)\left(\omega+i\Lambda_{-}\right)},
\\\label{g12-w}
g_{12}(\omega) &=& \frac{2i{\cal N}_o\lambda \left(2\bar n_{sp}+1\right)}
                {\left(\omega+i\Lambda_{+}\right)\left(\omega+i\Lambda_{-}\right)},
\\\label{g21-w}
g_{21}(\omega) &=& -\frac{2i\lambda \bar s_{z}}
                {\left(\omega+i\Lambda_{+}\right)\left(\omega+i\Lambda_{-}\right)},
\\\label{g22-w}            
g_{22}(\omega)&=& \frac{i\left(\omega-2\omega_{sp}+2i\gamma_{sp}\right)}
                    {\left(\omega+i\Lambda_{+}\right)\left(\omega+i\Lambda_{-}\right)}.
\end{eqnarray}

\bibliographystyle{prsty}
\bibliography{Manuscript-arXive-Final.bbl}

\end{document}